\documentclass[twocolumn]{aastex631}

\graphicspath{{./}}

\shorttitle{Mass of the Milky Way CGM}
\shortauthors{Ravi et al.}

\begin{document}

\title{Deep Synoptic Array science: a 50\,Mpc fast radio burst constrains the mass of the Milky Way circumgalactic medium}

\correspondingauthor{Vikram Ravi}
\email{vikram@caltech.edu}

\author{Vikram Ravi}
\affiliation{Cahill Center for Astronomy and Astrophysics, MC 249-17 California Institute of Technology, Pasadena CA 91125, USA.}
\affiliation{Owens Valley Radio Observatory, California Institute of Technology, Big Pine CA 93513, USA.}

\author{Morgan Catha}
\affiliation{Owens Valley Radio Observatory, California Institute of Technology, Big Pine CA 93513, USA.}

\author{Ge Chen}
\affiliation{Cahill Center for Astronomy and Astrophysics, MC 249-17 California Institute of Technology, Pasadena CA 91125, USA.}

\author{Liam Connor}
\affiliation{Cahill Center for Astronomy and Astrophysics, MC 249-17 California Institute of Technology, Pasadena CA 91125, USA.}

\author{James M. Cordes}
\affiliation{Department of Astronomy, Cornell University, Ithaca, NY 14853, USA}
\affiliation{Cornell Center for Astrophysics and Planetary Science and Department of Astronomy, Cornell University, Ithaca, NY 14853, USA}

\author{Jakob T. Faber}
\affiliation{Cahill Center for Astronomy and Astrophysics, MC 249-17 California Institute of Technology, Pasadena CA 91125, USA.}

\author{James W. Lamb}
\affiliation{Owens Valley Radio Observatory, California Institute of Technology, Big Pine CA 93513, USA.}

\author{Gregg Hallinan}
\affiliation{Cahill Center for Astronomy and Astrophysics, MC 249-17 California Institute of Technology, Pasadena CA 91125, USA.}
\affiliation{Owens Valley Radio Observatory, California Institute of Technology, Big Pine CA 93513, USA.}

\author{Charlie Harnach}
\affiliation{Owens Valley Radio Observatory, California Institute of Technology, Big Pine CA 93513, USA.}

\author{Greg Hellbourg}
\affiliation{Cahill Center for Astronomy and Astrophysics, MC 249-17 California Institute of Technology, Pasadena CA 91125, USA.}
\affiliation{Owens Valley Radio Observatory, California Institute of Technology, Big Pine CA 93513, USA.}

\author{Rick Hobbs}
\affiliation{Owens Valley Radio Observatory, California Institute of Technology, Big Pine CA 93513, USA.}

\author{David Hodge}
\affiliation{Cahill Center for Astronomy and Astrophysics, MC 249-17 California Institute of Technology, Pasadena CA 91125, USA.}

\author{Mark Hodges}
\affiliation{Owens Valley Radio Observatory, California Institute of Technology, Big Pine CA 93513, USA.}

\author{Casey Law}
\affiliation{Cahill Center for Astronomy and Astrophysics, MC 249-17 California Institute of Technology, Pasadena CA 91125, USA.}
\affiliation{Owens Valley Radio Observatory, California Institute of Technology, Big Pine CA 93513, USA.}

\author{Paul Rasmussen}
\affiliation{Owens Valley Radio Observatory, California Institute of Technology, Big Pine CA 93513, USA.}

\author{Kritti Sharma}
\affiliation{Cahill Center for Astronomy and Astrophysics, MC 249-17 California Institute of Technology, Pasadena CA 91125, USA.}

\author{Myles B. Sherman}
\affiliation{Cahill Center for Astronomy and Astrophysics, MC 249-17 California Institute of Technology, Pasadena CA 91125, USA.}

\author{Jun Shi}
\affiliation{Cahill Center for Astronomy and Astrophysics, MC 249-17 California Institute of Technology, Pasadena CA 91125, USA.}

\author{Dana Simard}
\affiliation{Cahill Center for Astronomy and Astrophysics, MC 249-17 California Institute of Technology, Pasadena CA 91125, USA.}

\author{Jean J. Somalwar}
\affiliation{Cahill Center for Astronomy and Astrophysics, MC 249-17 California Institute of Technology, Pasadena CA 91125, USA.}

\author{Reynier Squillace}
\affiliation{Cahill Center for Astronomy and Astrophysics, MC 249-17 California Institute of Technology, Pasadena CA 91125, USA.}

\author{Sander Weinreb}
\affiliation{Cahill Center for Astronomy and Astrophysics, MC 249-17 California Institute of Technology, Pasadena CA 91125, USA.}

\author{David P. Woody}
\affiliation{Owens Valley Radio Observatory, California Institute of Technology, Big Pine CA 93513, USA.}

\author{Nitika Yadlapalli}
\affiliation{Cahill Center for Astronomy and Astrophysics, MC 249-17 California Institute of Technology, Pasadena CA 91125, USA.}

\collaboration{200}{(The Deep Synoptic Array team)}

\begin{abstract}

We present the Deep Synoptic Array (DSA-110) discovery and interferometric localization of the so far non-repeating FRB\,20220319D. The FRB originates in a young, rapidly star-forming barred spiral galaxy, IRAS\,02044$+$7048, at just 50\,Mpc. Although the NE2001 and YMW16 models for the Galactic interstellar-medium (ISM) contribution to the DM of FRB\,20220319D exceed its total observed DM, we show that uncertainties in these models accommodate an extragalactic origin for the burst. We derive a conservative upper limit on the DM contributed by the circumgalactic medium (CGM) of the Milky Way: the limit is either 28.7\,pc\,cm$^{-3}$ and 47.3\,pc\,cm$^{-3}$, depending on which of two pulsars nearby on the sky to FRB\,20220319D is used to estimate the ISM DM. These limits both imply that the total Galactic CGM mass is $<10^{11}M_{\odot}$, and that the baryonic mass of the Milky Way is $\lesssim60\%$ of the cosmological average given the total halo mass. More stringent albeit less conservative constraints are possible when the DMs of pulsars in the distant globular cluster M53 are additionally considered. Although our constraints are sensitive to possible anisotropy in the CGM and to the assumed form of the radial-density profile, they are not subject to uncertainties in the chemical and thermal properties of the CGM. Our results strongly support scenarios commonly predicted by galaxy-formation simulations wherein feedback processes expel baryonic matter from the halos of galaxies like the Milky Way. 

\end{abstract}


\section{Introduction \label{sec:1}}

Galaxies like the Milky Way are embedded in a multiphase ($\sim10^{4}-10^{7}$\,K), highly ionized (hydrogen neutral fractions $\ll0.01$\%), kinematically complex, spatially clumpy and anisotropic circumgalactic medium \citep[CGM;][]{tpw17}. Likely extending beyond dark-matter halo virial radii $r_{\rm vir}$, the CGM may represent the dominant baryon component by mass within halos. The physical properties of the CGM of external galaxies, including density, temperature, kinematic structure, and chemical composition have long been probed by absorption-line measurements towards background objects. More recently, detections of thermal bremsstrahlung \citep[e.g.,][]{lbw+18} and the Sunyaev-Zeldovich effect around nearby galaxies \citep[e.g.,][]{bhq+22} provide independent constraints on the CGM density and temperature structure, potentially confirming the presence of extended coronae around galaxies. Evidence for extended gas reservoirs associated with nearby galaxies, either in the CGM or the intra-group medium, has also been observed in the dispersion measures (DMs) of background fast radio bursts \citep[FRBs;][]{cr22,wm22}. Galaxy-formation simulations highlight the dependence of CGM properties on feedback from stellar winds, supernovae, and AGN, as well as on the galaxy merger histories \citep[e.g.,][]{wso20,hfa+20,zpo+20,ftd+20,ads+21,rnp22}. The simulations predict varying degrees of anisotropy in the CGM of individual galaxies, and different total CGM masses for galaxies with comparable global characteristics. A ubiquitous prediction of simulations is that feedback processes determine the integrated CGM mass, $M_{\rm CGM}$. In general, CGM baryon fractions $f_{\rm CGM}=\frac{M_{\rm CGM}\Omega_{M}}{M_{\rm tot}\Omega_{b}}\lesssim0.5$ are predicted, where $M_{\rm tot}$ is the total (dark matter and baryonic) halo mass.

A halo of $\sim10^{6}$\,K gas at densities $\lesssim10^{-3}$\,cm$^{-3}$ surrounding the Milky Way was proposed by \citet{s56} as a solution to the problem of confining distant cold-gas clouds at high Galactic latitudes. Tentative evidence for such coronal gas had previously been found in observations of the diffuse radio-synchrotron background \citep[e.g.,][]{b56}. A theoretical basis for the origins of such a hot CGM around galaxies was developed in the 1970s \citep{ro77,wr78}, with the gas infalling onto dark-matter halos heated to virial temperatures of $\gtrsim10^{6}$\,K, possibly in shocks at the halo virial radii \citep[see also][]{bd03}. It would, however, be nearly two decades before further observational evidence for a hot Milky Way CGM was identified using observations of shadows in the soft X-ray background from high-latitude \citep{smh+91} and high-velocity \citep{hms+95} cold clouds, and in the spectral decomposition of the background into absorbed and unabsorbed components \citep{gna+92}. Today the hot CGM of the Milky Way is best traced through X-ray observations of OVII and OVIII diffuse emission, and absorption towards bright background sources. Although high Galactic latitude data are well fit by a spheroidal component with a mass of a few $\times10^{10}M_{\odot}$ \citep[e.g.,][]{bl07,gmk+12}, more recent models include a substantially less massive ($\sim0.1\%$), but denser disk-like component to explain the line observations \citep{lb17,niy+18,kkk+20,usk+22}. Structure in the hot Milky Way CGM is also observed in the form of the Fermi \citep{ssf10}, Parkes \citep{ccs+13} and eROSITA \citep{psb+20} bubbles. The neutral and warm ($\lesssim10^{5}$\,K) components of the Milky Way CGM, as primarily traced through observations of high-velocity clouds, are likely sub-dominant in mass to the hot component \citep[few $\times10^{8}M_{\odot}$;][]{ppj12}, but highlight the complexities of the disk-halo interface \citep[e.g.,][]{khr92,wrb+19,cbf22}. 

The mass of the Galactic CGM is uncertain. Here we assume $M_{\rm tot}=1.5\times10^{12}M_{\odot}$ \citep[following][]{pz19} and a total Galactic mass in stars and cold gas of $6\times10^{10}M_{\odot}$ \citep{d11}. Setting $f_{\rm CGM}=1$ and $\Omega_{b}/\Omega_{M}=0.188$ \citep{pc20} implies $M_{\rm CGM}=2.82\times10^{11}M_{\odot}$. Recent models for the distribution of OVII and OVIII emission and absorption result in $M_{\rm CGM}$ values between $3-4\times10^{10}M_{\odot}$ \citep{lb17}, $5.5-8.6\times10^{10}M_{\odot}$ \citep{kkk+20}, and $\sim1.2\times10^{11}M_{\odot}$ \citep{yt20}. \citet{fsm17} argue that a portion of the hot CGM is also traced by OVI absorption, as is observed in external galaxies \citep{ttw+11}, and find $M_{\rm CGM}=1.2\times10^{11}M_{\odot}$. Systematic effects that make the estimation of $M_{\rm CGM}$ difficult include the model uncertainties in foregrounds and in the spatial distribution and clumpiness of the gas, and the poorly constrained gas metallicity. An alternate measurement of $M_{\rm CGM}$ was found by \citet{sbb+15} by modeling the ram-pressure stripping of the Large Magellanic Cloud (LMC); a low value of $(2.7\pm1.4)\times10^{10}M_{\odot}$ was found. 

The DMs of FRBs and pulsars within the Milky Way halo offer a unique probe of the content of the Galactic CGM \citep[e.g.,][]{ab10,ppl20}. FRBs are transient radio emissions from distant extragalactic sources that, like the emissions from radio pulsars, are so short in duration that the dispersion in intervening plasma is evident as an arrival-time delay at lower frequencies:
\begin{equation}
    \Delta t(\nu) = \frac{q_{e}^{2}}{2\pi m_{e}c\nu^{2}}\int_{0}^{D}n_{e}(l)dl,
\end{equation}
where $q_{e}$ is the electron charge, $m_{e}$ is the electron mass, $\nu$ is the frequency, $D$ is the source distance, and $n_{e}(l)$ is the electron number density along the sightline. Under physical conditions of CGM gas, DMs quantify the column densities along the source sightlines. In all cases, however, the observed DMs of FRBs, ${\rm DM}_{\rm obs}$, are composed of a selection of additive components:
\begin{equation}
{\rm DM}_{\rm obs} = {\rm DM_{\rm ISM}} + {\rm DM_{\rm CGM}} + {\rm DM_{\rm IGM}} + {\rm DM_{\rm host}},
\end{equation}
where in this formulation ${\rm DM_{\rm ISM}}$ includes the warm ionized medium (WIM) in the Milky Way disk \citep[e.g.,][]{cl02,ymw17}, ${\rm DM_{\rm CGM}}$ is the DM contributed by the Galactic CGM, ${\rm DM_{\rm host}}$ is the DM associated with the FRB host-galaxy interstellar medium (ISM) and their halos, and ${\rm DM_{\rm IGM}}$ includes contributions from the intergalactic medium (IGM) and any intervening galaxy systems. For pulsars within the Milky Way halo, such as in the Magellanic Clouds and in several globular clusters, ${\rm DM_{\rm IGM}}=0$\,pc\,cm$^{-3}$. These expressions apply to the redshift $\sim0$ regime relevant to this paper. If the components of ${\rm DM}_{\rm obs}$ can be accurately modeled or bounded, constraints can be placed on ${\rm DM_{\rm CGM}}$. Modeling of the DMs of pulsars in the LMC led \citet{ab10} to find a very low $M_{\rm CGM}\sim1.5\times10^{10}M_{\odot}$ assuming a \citet{nfw97} profile (an NFW profile); it is likely that an overly large ${\rm DM_{\rm ISM}}$ was assumed, and the NFW profile is no longer favored for the Galactic CGM. Although most FRBs are too distant to allow for accurate modeling of the components of ${\rm DM}_{\rm obs}$, FRBs from nearby galaxies \citep{bkm+21,kmn+22} have begun to be used to place constraints on ${\rm DM_{\rm CGM}}$, assuming specific models for ${\rm DM_{\rm ISM}}$, that are in tension with some models for the Milky Way CGM. 

In this paper we present observations with the Deep Synoptic Array (DSA-110) of a new FRB (FRB\,20220319D) that motivates stringent new constraints on the mass of the Milky Way CGM. Remarkably, FRB\,20220319D was observed to have ${\rm DM}_{\rm obs}<{\rm DM_{\rm ISM}}$ for leading models of the warm-gas distribution in the Galactic disk, despite the unambiguous association we find with a galaxy at a distance of 50\,Mpc. We detail observations of the FRB in \S\ref{sec:2}. We consider the association with its host galaxy in \S\ref{sec:3}, wherein we show that the DM of FRB\,20220319D is consistent with an extragalactic origin given uncertainties in models for ${\rm DM_{\rm ISM}}$, using observations of the DMs of pulsars in globular clusters with accurate parallax distances. We present an analysis of the host environment of FRB\,20220319D in \S\ref{sec:4}. We then place conservative constraints on ${\rm DM_{\rm CGM}}$, and hence $M_{\rm CGM}$, using FRB\,20220319D and the distant high-latitude globular cluster M53, in \S\ref{sec:5}. We discuss the implications of our results in \S\ref{sec:6}, and conclude in \S\ref{sec:7}.

\section{DSA-110 observation of FRB\,20220319D \label{sec:2}} 

\begin{figure}
    \centering
    \includegraphics[width=0.48\textwidth,trim=0cm 9cm 0cm 0cm,clip]{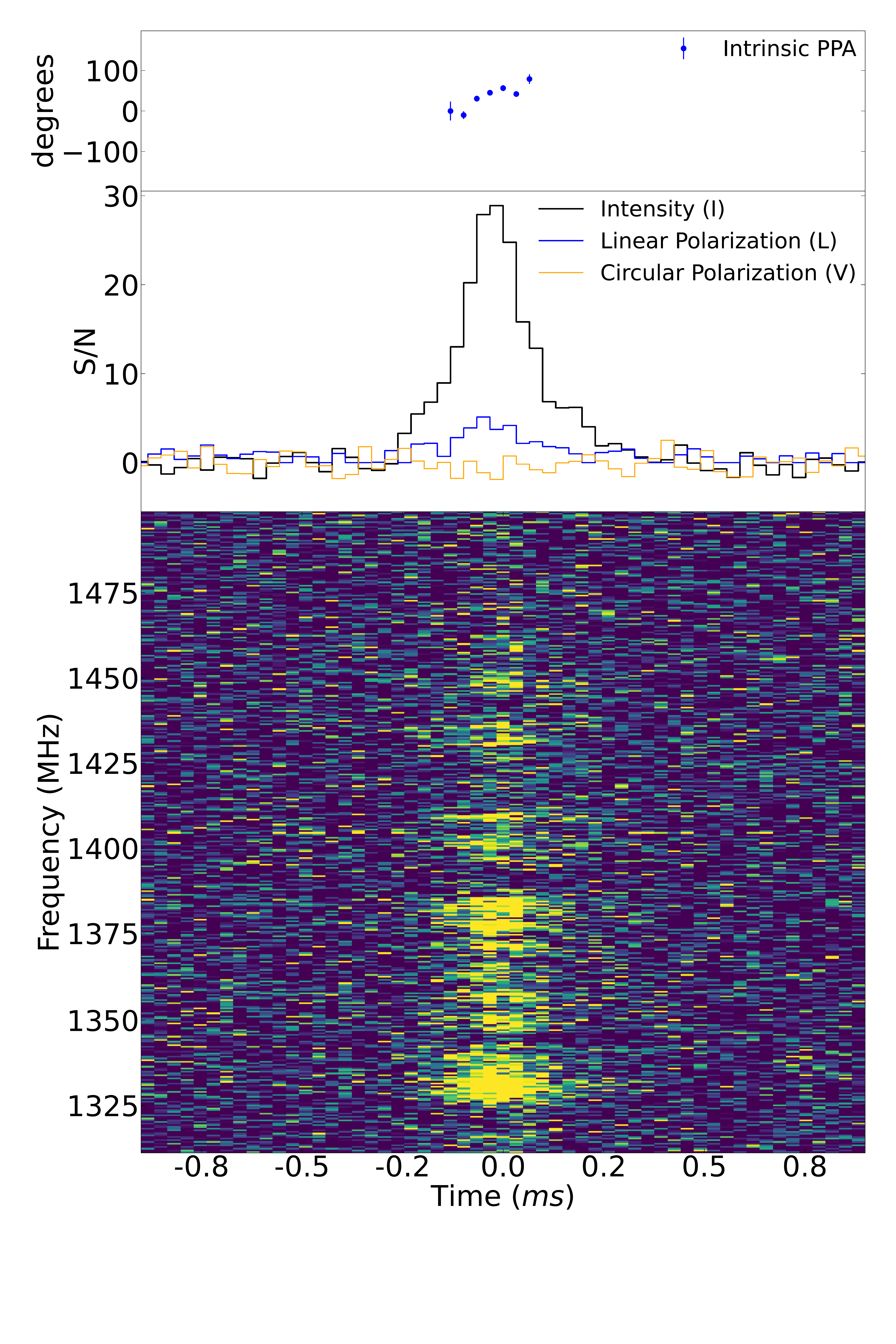}
    \caption{Dedispersed temporal profile and dynamic spectrum of FRB\,20220319D. A value of ${\rm DM}_{\rm obs}=110.95$\,pc\,cm$^{-3}$ was used. The top panel shows the estimated absolute position angle based on an approximate parallactic angle, the $\sim20^{\circ}$ error due to RM uncertainty is not included in the error bars. The middle panel shows temporal profiles in total intensity, linearly polarized intensity, and Stokes V, as labeled. The bottom panel shows the dynamic spectrum of the total-intensity data. The time resolution in the time-series and dynamic-spectrum plots is 32.768\,$\mu$s, and the temporal profile is in units of signal to noise ratio. The reference time is MJD 59657.93275696795, which was the burst arrival time at OVRO at 1530\,MHz. The polarized data on the burst was corrected for the measured RM of 50\,rad\,m$^{-2}$.}
    \label{fig:dynspec_pol}
\end{figure}

The DSA-110\footnote{\url{https://deepsynoptic.org}} is a radio interferometer hosted at the Owens Valley Radio Observatory (OVRO), purpose built for the discovery and arcsecond-localization of FRBs. A full description of the instrument will be presented in Ravi et al. (in prep.). During the observations discussed herein, the DSA-110 was configured as described in \citet{rcc+22}. Of particular note is that when FRB\,20220319D was observed, early in DSA-110 commissioning, all candidates at DMs in excess of 80\,pc\,cm$^{-3}$ were saved for further inspection, regardless of the expected DM through the Galactic disk.  

FRB\,20220319D was detected during standard commissioning observations, with an arrival time at OVRO at 1530\,MHz of MJD 59657.93275696795. During these observations, the DSA-110 was parked at a pointing-center declination of $+71.6^{\circ}$. The DSA-110 reference position is $-118.283^{\circ}$ longitude, $+37.2334^{\circ}$ latitude. FRB\,20220319D was detected with a DM of 110.96\,pc\,cm$^{-3}$, and a signal to noise ratio (S/N) of 41.7. To derive optimized burst parameters and polarization properties, we coherently combined voltage data from the 48 core antennas towards the detection-beam direction. The resulting temporal profile and calibrated dynamic spectrum, produced with incoherent dedispersion at the native time and frequency resolutions, are shown in Figure~\ref{fig:dynspec_pol}, and optimized parameters are given in Table~\ref{tab:frb}. The polarization analysis was done following procedures described in \citet{rcc+22}. The matched-filter total-intensity S/N estimate is 79. 

The burst exhibits a narrow temporal profile with a strongly modulated spectrum. The DM smearing timescale ranges between 8--13\,$\mu$s within the DSA-110 band, and so the data resolve the temporal structure of the burst. Scaling the system-equivalent flux density of the DSA-110 by the primary-beam attenuation at the burst position, we derive a fluence of $11.2\pm0.2$\,Jy\,ms. Following techniques described in \citet{cvo+20}, we derive a spectral decorrelation bandwidth, $\nu_{d}=3.6\pm0.2$\,MHz, defined as the $1/e$ scale of the autocorrelation function of the spectrum. The burst spectro-temporal morphology is consistent with the population of so far non-repeating FRBs \citep{pgk+21}. We derive a low fractional polarization of just 16\% (Figure~\ref{fig:dynspec_pol}), with no significant circular polarization. A Faraday rotation measure (RM) of $50\pm15$\,rad\,m$^{-2}$ was detected. The error was estimated by simulating the recovery of RMs with the same linear-polarization S/N as FRB\,20220319D (Sherman et al., in prep.). 

\begin{figure*}
    \centering
    \includegraphics[width=0.95\textwidth]{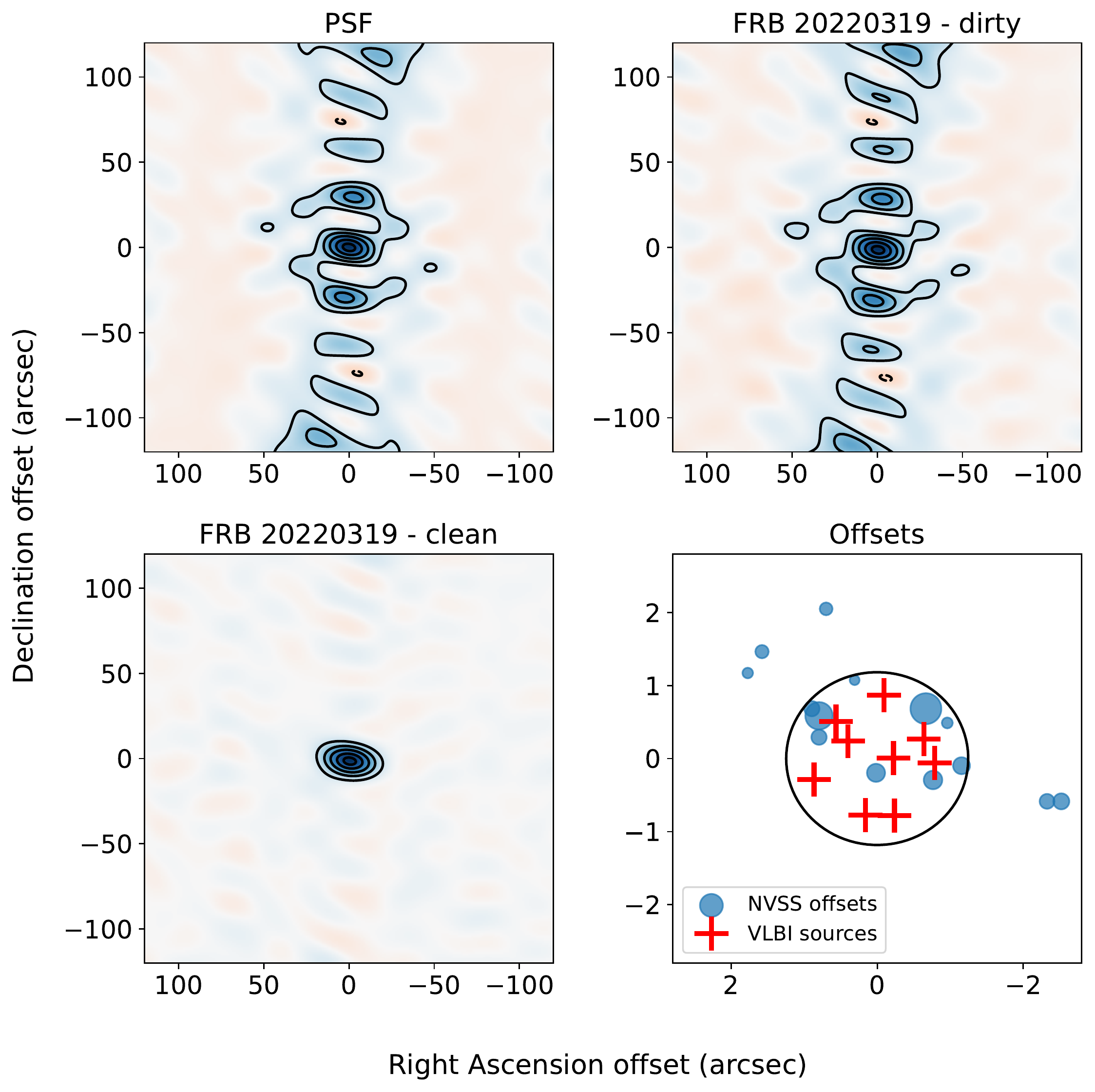}
    \caption{DSA-110 localization of FRB\,20220319D. \textit{Top left:} Point-spread function (PSF) of the DSA-110 for the observation of FRB\,20220319D, assuming a flat-spectrum source. \textit{Top right:} Dirty image of FRB\,20220319D, with no deconvolution applied. \textit{Bottom left:} Deconvolved image of FRB\,20220319D. The synthesized beam is represented by an ellipse with half-power diameters of 26.9\arcsec$\times$15.6\arcsec, at a position angle of 80$^{\circ}$. Briggs weighting was used with a `robust' parameter of 0.5, in order to partially suppress PSF sidelobes. Baselines shorter than $200\lambda$ were excluded from the analysis because they were affected by spurious cross-coupling. Contours in the preceding three panels are at $-0.4$, $-0.2$ (dashed), $0.2$, $0.4$, $0.6$, $0.8$ and $0.9$ (solid) of the peak intensity. The images are centered on the coordinates (R.A. J2000, decl. J2000) = (02:08:42.7, $+$71:02:06.9). \textit{Bottom right:} Offsets of known astronomical sources from their true positions as measured by the DSA-110. The blue disks show offsets of sources from cataloged positions in the NRAO VLA Sky Survey \citep[NVSS;][]{ccg+98} in a 5\,min DSA-110 image formed from data taken at a time centered on the burst detection. Only sources with cataloged flux densities $>50$\,mJy and measured major-axis diameters $<20$\arcsec, within 2\,deg of the pointing center, were considered. The symbol size is proportional to flux density. The red crosses show measured offsets of nine RFC calibrator sources (see text for details) observed immediately preceding and after the burst. The ellipse indicates the  90\% confidence containment region for FRB\,20220319D, derived from the RFC calibrators.}
    \label{fig:loc}
\end{figure*}

The interferometric localization of FRB\,20220319D was derived following procedures largely described in \citet{rcc+22}. Correlation products for all baselines formed from the 63 functioning antennas of the DSA-110 at the time of detection were integrated over 262.144$\mu$s centered on the burst arrival time. Bandpass calibration was done using a 10-min observation of 3C309.1 observed 11\,hr prior to the burst detection. Complex gain calibration at the time of the burst detection was derived using an NRAO VLA Sky Survey \citep[NVSS;][]{ccg+98} model and 5\,min of visibility data. We show images of the point-spread function (PSF; a flat spectrum was assumed), and the pre- and post-deconvolution images of FRB\,20220319D in Figure~\ref{fig:loc}. Briggs weighting with a `robust' parameter of 0.5 was used to suppress sidelobes, given the high S/N of FRB\,20220319D, yielding a PSF FWHM of $26.9\arcsec\times15.6\arcsec$. The position of FRB\,20220319D, given in Table~\ref{tab:frb}, was derived by fitting an elliptical Gaussian to the deconvolved image of FRB\,20220319D. 

We derived the uncertainty in the FRB position using data on nine compact bright sources from the Radio Fundamental Catalog (RFC; rfc\_2022c), obtained within 2\,hr of the burst observation. A fit to a linear trend in R.A. and Decl. was used to derive arcsecond-level astrometric corrections, and the corresponding uncertainties. These uncertainties (0.52\arcsec~in R.A. and 0.5\arcsec~in Decl.) were then added in quadrature to the statistical uncertainty in the fitted burst position (0.29\arcsec~in R.A. and 0.17\arcsec~in Decl.) to derive a final 90\% confidence containment ellipse with semi-axes of 1.25\arcsec~in R.A. and 1.18\arcsec~in Decl. This ellipse is shown in the bottom-left panel of Figure~\ref{fig:loc}, together with the detrended position offsets of the RFC sources. We also assessed the quality of the in-field calibration by checking the positions of bright ($>50$\,mJy) compact ($<20$\arcsec) NVSS sources within the primary-beam FWHM against NVSS catalog positions. These offsets (with the RFC corrections applied) are also shown in Figure~\ref{fig:loc}. A more detailed analysis of the DSA-110 localization accuracy will be presented in Ravi et al. (in prep.). 

\begin{deluxetable}{cc}
\tablecaption{Properties of FRB\,20220319D.\label{tab:frb}}
\tablehead{
\colhead{Parameter} & \colhead{Value}}
\startdata
Arrival time (MJD)\tablenotemark{a} & 59657.93275696795 \\
DM (pc\,cm$^{-3}$) & 110.95(1) \\
Full-width half-maximum (ms) & 0.16(1) \\
Fluence (Jy\,ms) & 11.2(2) \\
Spectral energy (erg\,Hz$^{-1}$) & 3.3(1)$\times10^{28}$ \\
$L/I$\tablenotemark{b} & 0.16(3) \\
$|V/I|$\tablenotemark{b} & 0.04(3) \\
RM (rad\,m$^{-2}$) & 50(15) \\
$\nu_{d}$ (MHz)\tablenotemark{c} & 3.6(2) \\
R.A. (J2000) &  02:08:42.7(1) \\
Decl. (J2000) & $+$71:02:06.9(6) 
\enddata
\tablenotetext{a}{Arrival time at OVRO at 1530\,MHz.}
\tablenotetext{b}{$L/I$ is the fraction of linearly polarized fluence, and $|V/I|$ is the absolute value of the fraction of circularly polarized fluence.}
\tablenotetext{c}{$\nu_{d}$ is the characteristic spectral decorrelation bandwidth.}
\end{deluxetable}

\section{Association with IRAS\,02044$+$7048, and uncertainties in ${\rm DM}_{\rm ISM}$ \label{sec:3}}

\begin{figure}
    \centering
    \includegraphics[width=0.48\textwidth]{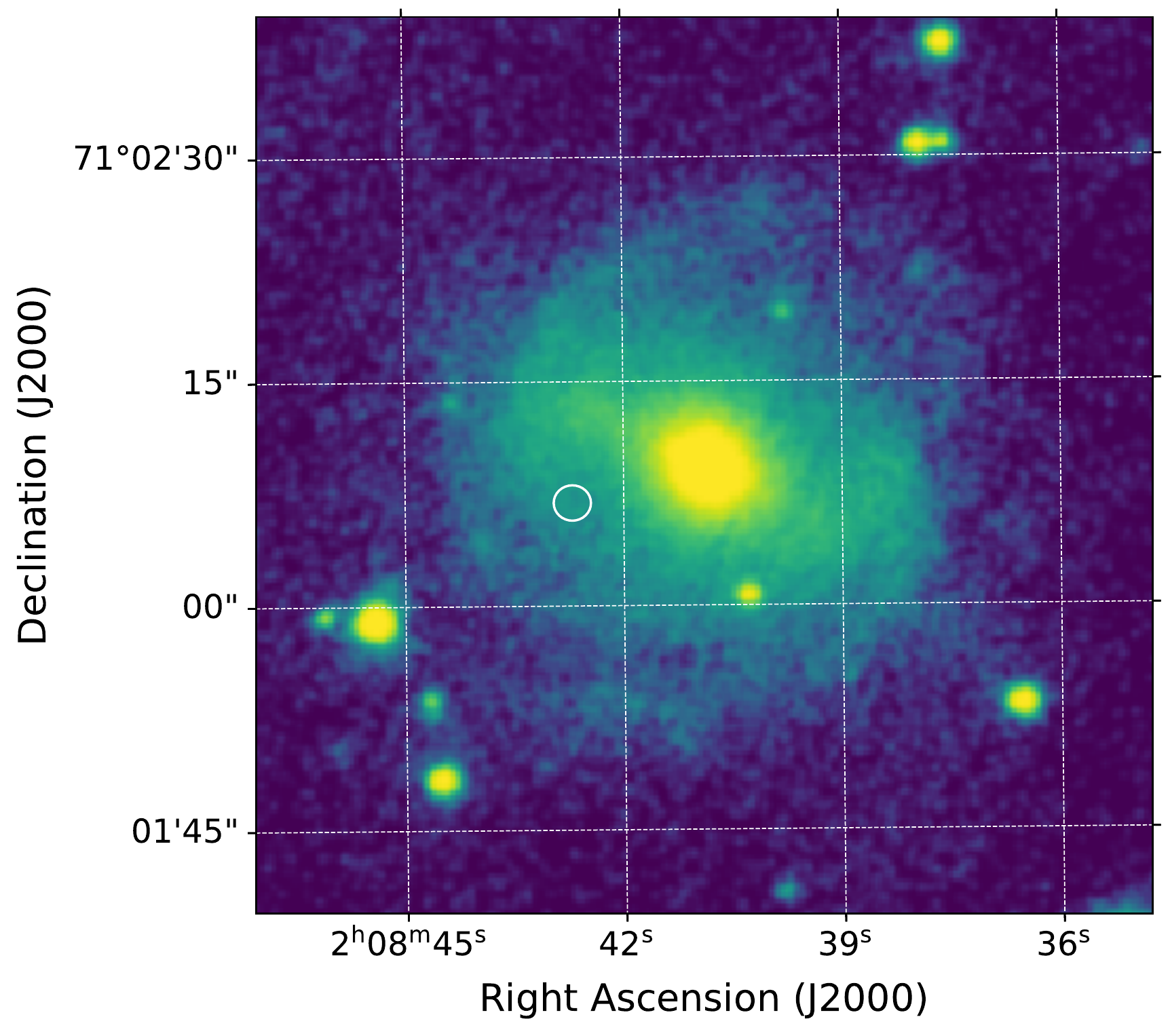}
    \caption{Pan-STARRS $i$-band image of IRAS\,02044$+$7048, with the $3.7\times2.2$\,arcsec 90\% confidence localization region of FRB\,20220319D indicated as a white ellipse.}
    \label{fig:image}
\end{figure}

The 90\% confidence localization region of FRB\,20220319D is shown in Figure~\ref{fig:image} overlaid on a Pan-STARRS1 (PS1) $i$-band image. The galaxy coincident with the FRB location is cataloged in the NASA Extragalactic Database as IRAS\,02044$+$7048, with a spectroscopic redshift of 0.011 indicating a distance of approximately 50\,Mpc. Using observations described below, we derive a spectroscopic redshift of $0.0111\pm0.0004$, indicating a luminosity distance of 49.6\,Mpc. An isophotal fit to the PS1 $i$-band image indicates an effective radius of $2.7\pm0.2$\,kpc, and the FRB is offset by $2.3\pm0.3$\,kpc. The isophotal analysis also indicates an inclination of $23\pm3$\,deg. A barred spiral morphology, with a classification of SBa, is evident from the image. 

Under the assumption that FRB\,20220319D is extragalactic, the spatial association between FRB\,20220319D and IRAS\,02044$+$7048 is secure. It is unlikely that FRB\,20220319D lies significantly beyond IRAS\,02044$+$7048; the morphology of the galaxy indicates the presence of a warm-ISM phase, and scattering in this ISM would likely cause temporal broadening of several hundred milliseconds at our observing frequencies \citep{coc22,occ+22}. The Gravitational Wave Galaxy Catalog \citep[GWGC;][]{wdd11} lists 15869 galaxies at distances less than 50\,Mpc, subtending a total of 4\,deg$^{2}$. Thus the chance-association probability of FRB\,20220319D and a galaxy at $<50$\,Mpc is $\lesssim10^{-4}$. 

The extragalactic nature of FRB\,20220319D is called into question by its DM of $110.95$\,pc\,cm$^{-3}$. The predicted DM for extragalactic objects along its $l=129.2^{\circ}$, $b=+9.1^{\circ}$ sightline is 132.9\,pc\,cm$^{-3}$ according to the NE2001 model for the Galactic ionized ISM distribution \citep{cl02}, and 187.7\,pc\,cm$^{-3}$ according to the YMW16 model \citep{ymw17}. The RM detection for FRB\,20220319D of $50\pm15$\,rad\,m$^{-2}$ is, however, marginally consistent with being in excess of the expected Milky Way contribution along the burst sightline. A recent model for the Galactic RM foreground along this sightline \citep{hab+22} indicates an expected Galactic RM of $-14\pm18$\,rad\,m$^{-2}$ towards FRB\,20220319D. We cannot draw conclusions from the measured spectral decorrelation bandwidth of $\nu_{d}=3.6\pm0.2$\,MHz, which is far in excess of the  0.2\,MHz decorrelation bandwidth predicted by NE2001 due to scattering in the Milky Way ISM, because the origins of the spectral structure cannot easily be determined. 

We now analyze the performance of the NE2001 and YMW16 models with the aim of determining whether the model uncertainties can accommodate an extragalactic origin for FRB\,20220319D. Both model the distribution of the WIM \citep{d11} in the Galactic disk using observations of radio-wave propagation, with a dominant thick-disk component encapsulating several overdensities and voids motivated by multiwavelength data. The YMW16 model is fit to pulsars with independent DM and distance estimates, whereas the NE2001 model is additionally fit to pulsar and extragalactic radio-source scattering measurements. For low-latitude extragalactic sightlines, the radial extent of the thick disk is critical in determining ${\rm DM_{\rm ISM}}$. In NE2001, a cutoff at a radius of 20\,kpc is motivated by observations of HII regions in other galaxies and scattering of extragalactic sources. Surveys of Galactic HII regions on the other hand motivate the cutoff of 15\,kpc in YMW16. Neither cutoff is estimated in the NE2001 or YMW16 fits; rather they are fixed model choices that do not impact the match between the models and the data under consideration. The functional forms for the WIM distribution in the thick disk also differ, with NE2001 incorporating an oblate spheroid, and YMW16 incorporating a plane-parallel slab that results in significant deviations from NE2001 at low latitudes in the second and third quadrants \citep[e.g.,][]{pfd21}. High-latitude extragalactic sightlines are likely to be better modeled because the scale height of the thick disk is a critical fitted parameter. However, density fluctuations caused by turbulence in the WIM, together with clumps and voids, complicate our picture even at high latitudes \citep{occ20}. Large-scale WIM inhomogeneities at the disk-halo interface such as Galactic chimneys \citep[e.g.,][]{khr92,ntd96} can result in significant departures from spatially smooth models for ${\rm DM_{\rm ISM}}$. 

We first note that the recent PSR$\pi$ sample of pulsar parallax measurements \citep{dgb+19} reveals discrepancies between the NE2001 and YMW16 models and measured pulsar distances at low latitudes in the second quadrant, where FRB\,20220319D appears on the sky. In general, known errors in NE2001 and YMW16 exhibit significant spatial correlation \citep[e.g.,][]{pfd21}, motivating the consideration of sources along sightlines similar to FRB\,20220319D. Table~\ref{tab:psr} lists the DMs of the three nearest pulsars in the sample to FRB\,20220319D, together with the predicted DMs given the measured distances. The YMW16 model significantly overpredicts all three DMs, and although NE2001 is more accurate for the modestly distant pulsars likely in the Cygnus-Orion arm, it also overpredicts the DM of the more distant PSR\,J0406$+$6138. 
 
\begin{deluxetable*}{cccccccc}
\tablecaption{Measured and predicted DMs of PSR$\pi$ pulsars near FRB\,20220319D. \label{tab:psr}}
\tablehead{
\colhead{Pulsar} & \colhead{$l$ (deg.)} & \colhead{$b$ (deg.)} & \colhead{FRB offset (deg)} &  \colhead{Distance (kpc)} & \colhead{DM\tablenotemark{a}} & \colhead{NE2001\tablenotemark{a}} & \colhead{YMW16\tablenotemark{a}}}
\startdata
PSR\,J0147$+$5922 & 130.1 & $-$2.7 & 12 & $2.02^{+0.46}_{-0.16}$ & 40.1 & 31.8 & 79.0 \\
PSR\,J0157$+$6212 & 130.6 & 0.3 & 9 & $1.80^{+0.08}_{-0.12}$ &  30.2 & 31.5 & 59.3 \\
PSR\,J0406$+$6138 & 144.0 & 7.0 & 15 & $4.58^{+1.63}_{-0.87}$ & 65.3 & 123.7 & 141.7 
\enddata
\tablenotetext{a}{All DMs are given in units of pc\,cm$^{-3}$.}
\end{deluxetable*}

\begin{figure*}
    \centering
    \includegraphics[width=0.85\textwidth]{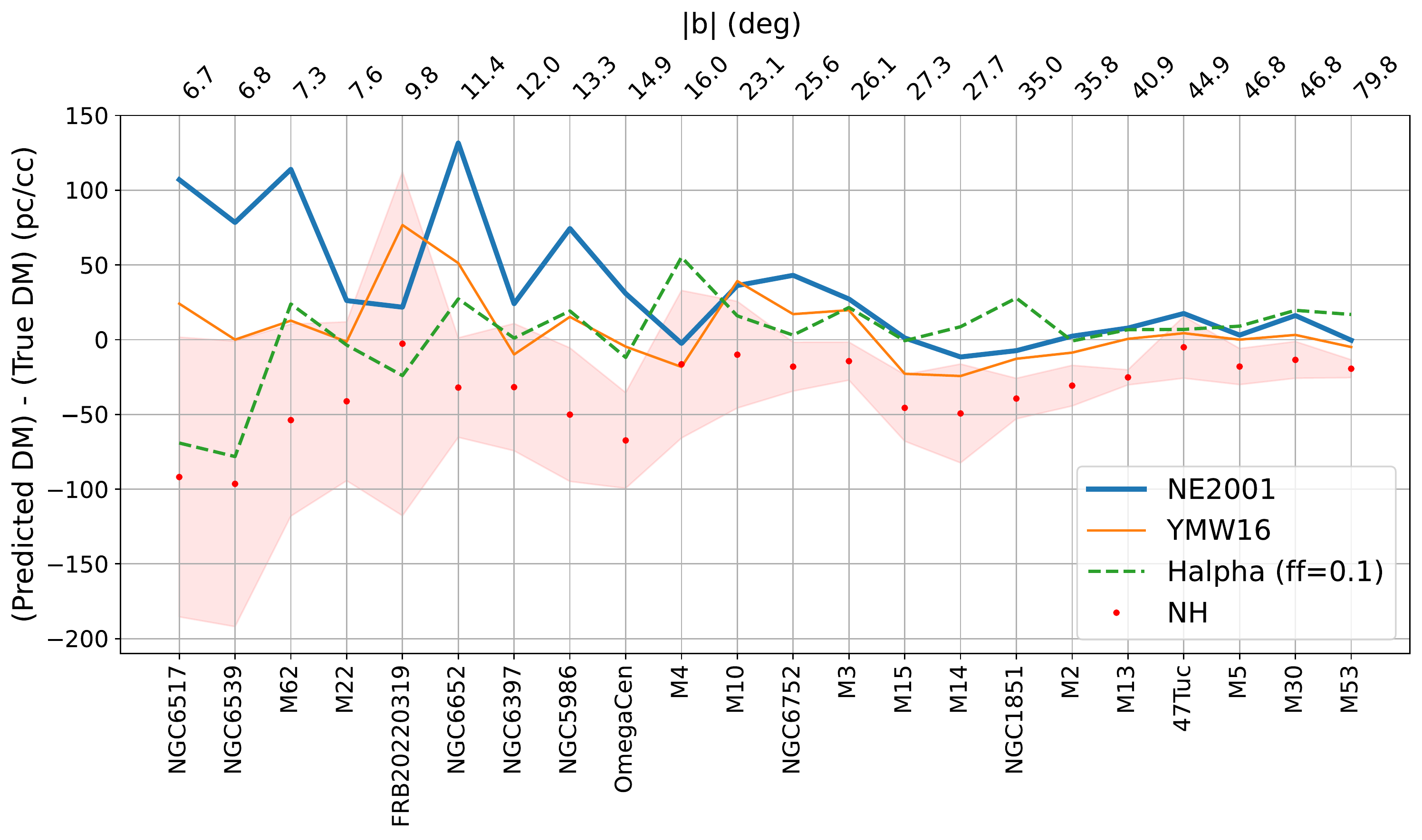}
    \caption{Difference between the predicted DM and true DM of all globular clusters that host pulsars, sorted by increasing $|b|$. Predictions are made for each globular cluster given its distance and location on the sky, using four different models for the warm ISM of the Milky Way. The two literature models are NE2001 (blue solid thick curve)  and YMW16 (orange solid thin curve). We also convert the measured H${\alpha}$ emission measures (green dashed curve) and HI columns (red dots and shaded region) along each sightline to DM predictions using methods described in the text. We also include predictions for FRB\,20220319D assuming a location beyond the Galactic disk.}
    \label{fig:globs}
\end{figure*}

A more detailed analysis of the performance of the models for sightlines towards distant objects is enabled by the recent convergence in the distances to globular clusters hosting pulsars. We obtain accurate distance estimates (less than few-percent errors) to all 21 globular clusters with associated pulsars \citep{mht+05} at $|b|>5^{\circ}$ from the compilation of \citet{bv21}, primarily based on \textit{Gaia} parallaxes \citep{vb21}. Clusters at lower latitudes are not considered because their DMs are influenced by the thin disk and other components in the Galactic plane. For each cluster we compare the difference between the DMs predicted by the NE2001 and YMW16 models for the cluster distances, and the true cluster DMs defined as the mean DMs of the associated pulsars. The results are shown in Figure~\ref{fig:globs}; FRB\,20220319D is also included with the predicted DM set to ${\rm DM_{\rm ISM}}$. The errors in NE2001 at low latitudes can clearly accommodate an extragalactic origin for FRB\,20220319D. YMW16 generally performs better than NE2001 for the low-latitude clusters, but these are nearly all in the first and fourth quadrants (with NGC\,1851 as the only exception), and so perhaps not representative of errors along sightlines closer to the Galactic anticenter.

As an additional check of the Galactic sightline towards FRB\,20220319D, we also consider independent measures of ${\rm DM_{\rm ISM}}$. First, we use the relation between the neutral hydrogen column density and pulsar DMs derived by \citet{hnk13} to estimate the DMs towards FRB\,20220319D and each globular cluster, assuming no HI gas beyond the clusters. We obtain HI column densities from the HI4PI data \citep{hi4pi}. The results are shown in Figure~\ref{fig:globs}, including errors representing intrinsic scatter in the relation. The HI column densities result in underestimates of the DMs towards low-latitude clusters, possibly reflecting the increased ionization of HI gas towards the central regions of the Galaxy \citep[e.g.,][]{mr05}. The HI-predicted ${\rm DM_{\rm ISM}}$ for FRB\,20220319D is consistent with an extragalactic origin, although the uncertainties are large. Second, we estimate the DMs towards FRB\,20220319D and each globular cluster using H${\rm \alpha}$ emission measures (EMs) derived from the Wisconsin H${\rm \alpha}$ Mapper (WHAM) all sky maps \citep{hrt+03}. We follow the analysis of \citet{bmm06}, wherein the photon fluxes, $F_{\rm H \alpha}$ (in units of Rayleighs), integrated over the range of Galactic velocities are converted to DMs using the following relations:
\begin{eqnarray}
{\rm EM} &=& 2.25F_{\rm H \alpha}e^{2.2E(B-V)}\,{\rm cm^{-6}\,pc} \\
{\rm DM} &=&  ({\rm EM} \times D \times \mathcal{F})^{1/2}. \\
\mathcal{F} &=& \frac{f}{\zeta (1+\epsilon^{2})}
\end{eqnarray}
The expression for $\mathcal{F}$ \citep[e.g.,][and references therein]{occ20} encodes an ionized cloudlet model, wherein the WIM is composed of cloudlets with a volume filling factor $f$,  $\zeta=\langle n_{e}^{2}\rangle/\langle n_{e}\rangle/^{2}\sim2$ captures the cloud to cloud variation, and $\epsilon^{2}\lesssim1$ is the density variance internal to the cloudlets. Further, $E(B-V)$ is a measure of the interstellar redenning along a given sightline, and $D$ is the distance through the WIM. We have assumed a uniform WIM electron temperature of 8000\,K, and equivalence between the line-of-sight and volume filling factors. We find that for the cluster sightlines the H${\rm \alpha}$-based DM estimates perform remarkably well for a nominal value $\mathcal{F}=0.1$ (i.e., $f\gtrsim0.4$), comparably to YMW16 and better than NE2001 for low-latitude sightlines. This is surprising given previous results for pulsar DMs \citep[e.g.,][]{s12}, but may be explained by globular clusters generally lying beyond the outer extent of the WIM, obviating the need for distance corrections to the WHAM EMs. For the FRB\,20220319D sightline, the H${\rm \alpha}$ flux implies ${\rm DM_{\rm ISM}}=89$\,pc\,cm$^{-3}$, which is 22\,pc\,cm$^{-3}$ lower than the measured DM. 

We conclude that models for the DM towards FRB\,20220319D contributed by the Galactic WIM are consistent with an extragalactic origin. This secures the association of the burst with the galaxy IRAS\,02044$+$7048. Consistent with the original presentations of the NE2001 and YMW16 models for the WIM distribution, we confirm that there are significant model uncertainties at low Galactic latitudes. We used a large sample of globular-cluster parallax distances, which have DM estimates from associated pulsars, to demonstrate that an alternative predictor of ${\rm DM_{\rm ISM}}$ based on the total Galactic H${\rm \alpha}$ flux can provide a useful check on the aforementioned models. In order to estimate ${\rm DM_{\rm ISM}}$ towards FRB\,20220319D  we consider pulsars that are nearby on the sky and at a similar Galactic latitude. The two nearest pulsars within 1\,deg of the burst in Galactic latitude, PSR\,J0231$+$7026 (total separation of 1.9\,deg) and PSR\,J0325$+$6744 (total separation of 7.4\,deg), have DMs of 46.7\,pc\,cm$^{-3}$ and 65.3\,pc\,cm$^{-3}$, and we consider these as conservative and less conservative estimates respectively of ${\rm DM_{\rm ISM}}$ along the burst sightline. We note that the predicted NE2001 and YMW16 ${\rm DM_{\rm ISM}}$ values along the sightlines towards these pulsars are lower than along the burst sightline.

\section{The host environment of FRB\,20220319D \label{sec:4}}

\begin{deluxetable}{cc}
\tablecaption{Observed and derived parameters of the host galaxy of FRB\,20220319D, IRAS\,02044$+$7048. $1\sigma$ errors in the last significant figures are given in parentheses. \label{tab:gal}}
\tablehead{
\colhead{Parameter} & \colhead{Value}}
\startdata
Redshift &  0.0111(4) \\
Luminosity distance (Mpc) & 49.6  \\
Effective radius (kpc) & 2.7(2) \\
FRB projected offset (kpc) & 2.3(3) \\
$\log M_{*}$ ($M_{\odot}$) & 9.93(7) \\
$\log Z$\tablenotemark{a} & $0.1(0.2)$  \\
Internal $A_{V}$\tablenotemark{b} & $0.2^{+0.2}_{-0.1}$ \\
SFR ($M_{\odot}$\,yr$^{-1}$)\tablenotemark{c} & 1.8(0.7) \\
Inclination (deg) & 23(3) 
\enddata
\tablenotetext{a}{Metallicity with respect to solar.}
\tablenotetext{b}{$V$-band extinction corresponding to a uniform dust slab.}
\tablenotetext{c}{Averaged over the past 100\,Myr.}
\end{deluxetable}

\begin{figure}
    \centering
    \includegraphics[width=0.48\textwidth]{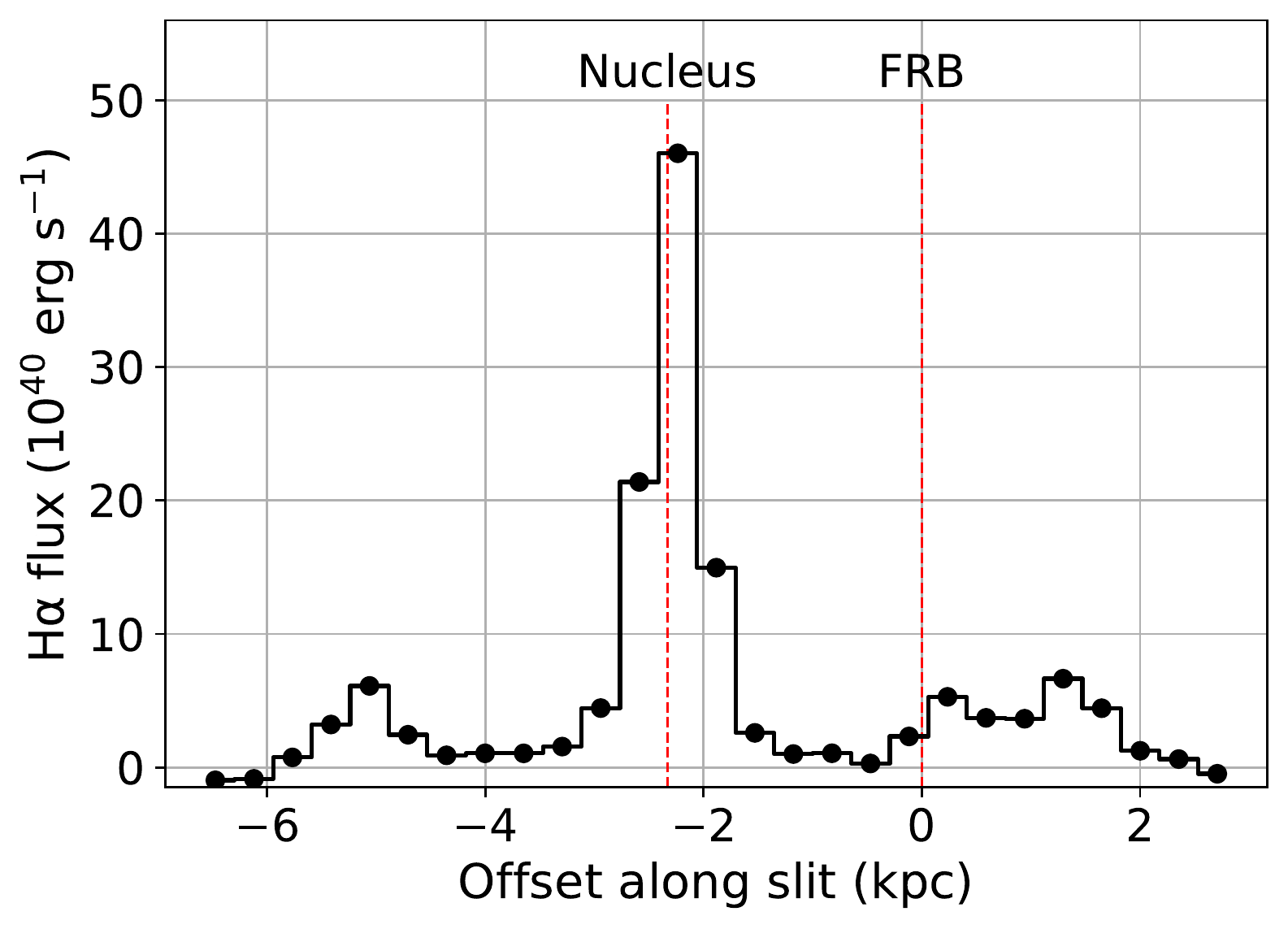}
    \caption{Total H${\rm \alpha}$ flux observed at different locations within our longslit spectroscopic observations of IRAS\,02044$+$7048. Measurements were obtained at 1.5\arcsec~intervals along a 1.5\arcsec~longslit oriented at a position angle of 104$^{\circ}$, which covered both the FRB position and the center of the galaxy. The locations of the galaxy nucleus and FRB\,20220319D are indicated in the figure.}
    \label{fig:spec}
\end{figure}

\begin{figure*}
    \centering
    \includegraphics[width=0.45\textwidth]{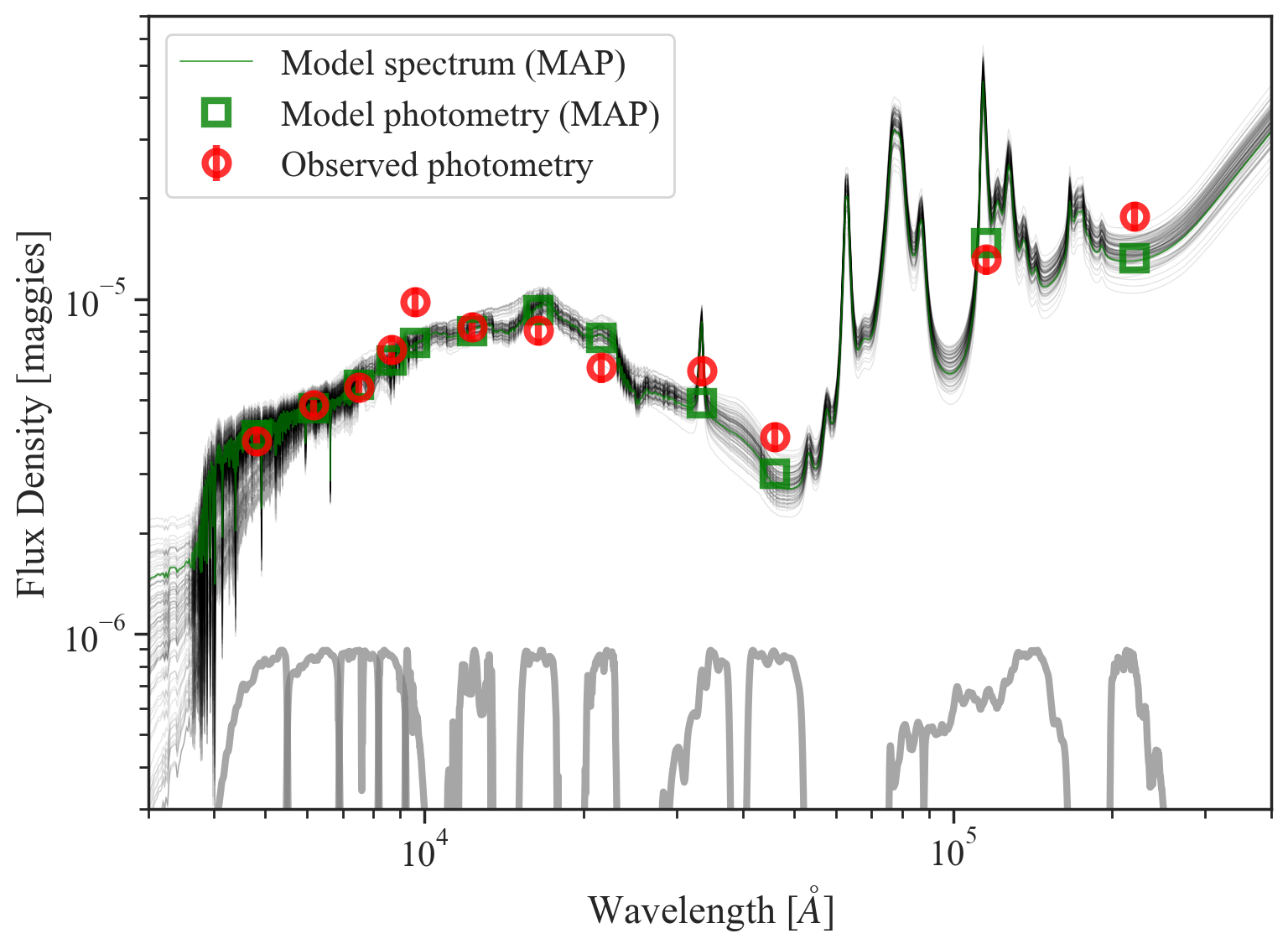}
    \includegraphics[width=0.45\textwidth]{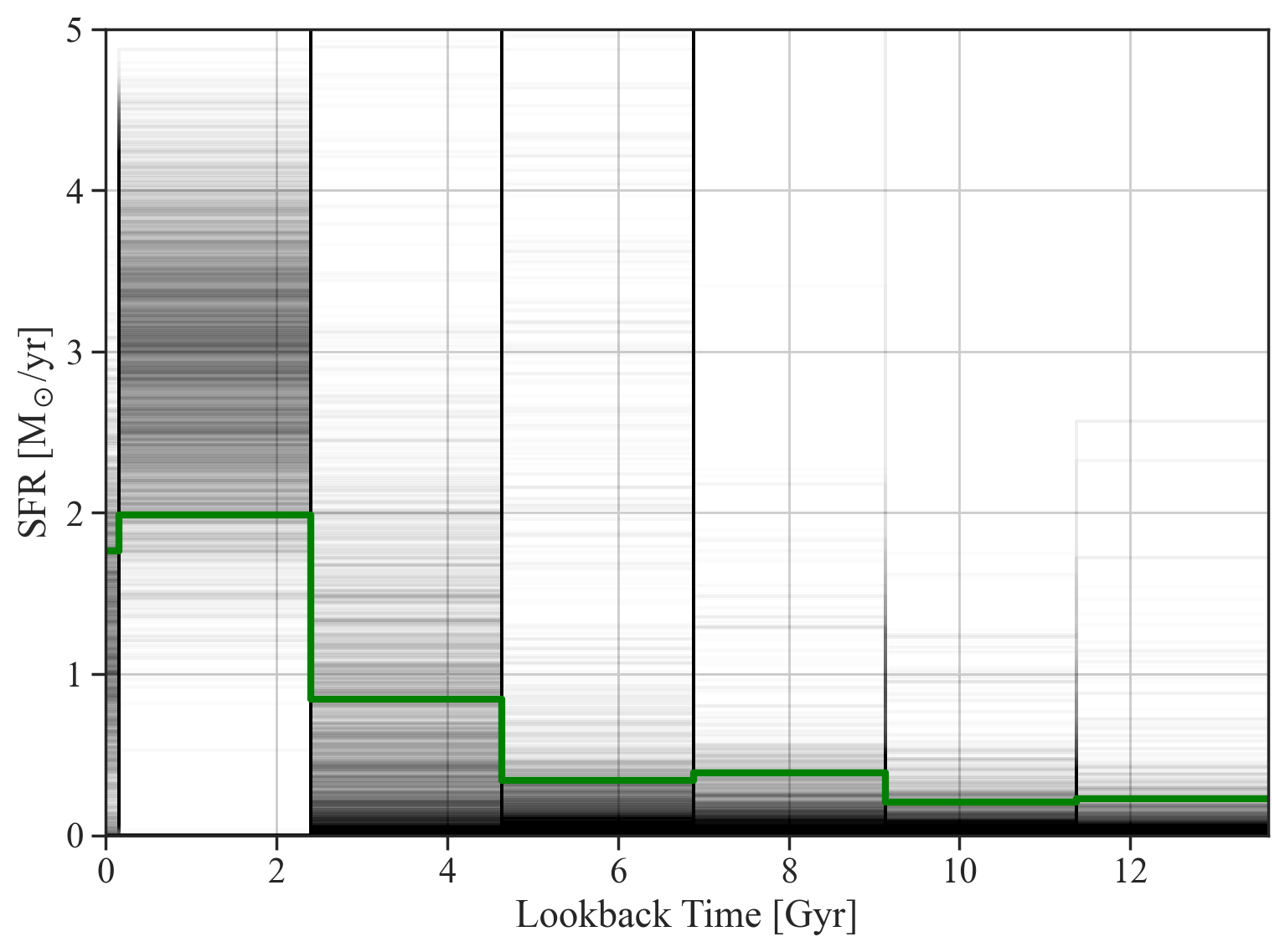}
    \caption{\textit{Left:} \texttt{Prospector} fit to the spectral energy distribution (SED) of the host galaxy of FRB\,20220319D, IRAS\,02044$+$7048. The SED measurements from archival PanSTARRS, 2MASS, and WISE data are shown in red, together with representative 10\% errors. Filter transmission curves are shown in grey. The maximum aposteriori probability (MAP) model photometry is shown in green, together with the MAP spectrum. 100 spectra generated using random draws from the posterior distributions of the model parameters are shown in black. \textit{Right:} Measured SFH of IRAS\,02044$+$7048 (green). The results from 100 random draws from the posterior distributions are shown in black.}
    \label{fig:sed}
\end{figure*}

We obtained long-slit optical spectroscopy of IRAS\,02044$+$7048 using the Double Spectrograph \citep[DBSP;][]{og82} mounted on the Palomar 200-inch Hale telescope on 2022 June 01 (UT). The observations were undertaken under conditions of $1.2$\arcsec~seeing, at an airmass of 1.9, and used a 1.5\arcsec~slit positioned to include both the FRB location and the galaxy nucleus at a position angle of 104$^{\circ}$. Two 600\,s exposures were obtained with the D55 dichroic and the 316/7500 grating on the red arm; we only considered data from the red arm of the spectrograph given the significant extinction along this sightline \citep[$A_{V}=2.209$;][]{sf11}. The two-dimensional spectral data were reduced according to standard procedures using a \texttt{PypeIt}-based pipeline \citep{phw+20}. We then defined the trace function using observations of the standard star Feige\,34, and extracted spectra in 1.5\arcsec~windows along the slit. Feige\,34 was also used for flux calibration, and all spectra were corrected for extinction according to the \citet{fm07} extinction curve.

We detected H${\rm \alpha}$ emission at several locations along the slit, as shown in Figure~\ref{fig:spec}. The strongest H${\rm \alpha}$ emission is observed at the galaxy core, from which we also detect lines corresponding to [NII] and [SII]. These lines were together used to derive a redshift of $0.0111\pm0.0004$ for IRAS\,02044$+$7048. Although we do not detect H${\rm \beta}$ or [OIII], the ratio $\log {\rm [NII] / H\alpha} = -0.23\pm0.02$ indicates a $<10\%$ chance of the nuclear ionization being purely due to star formation activity \citep{hfs97,kgk+06}. There is evidence of H${\rm \alpha}$ emission from spiral-arm features on either side of the nucleus; these features are too extended to represent HII regions. The FRB appears associated with the eastern arm, although not at the center of its H${\rm \alpha}$ radial profile. No additional evidence for or against an arm association can be gleaned from the residuals of a fit of the IRAS\,02044$+$7048 image in Figure~\ref{fig:image} to a two-dimensional Sersic profile. 

In order to determine global properties of IRAS\,02044$+$7048, we derived and modeled its spectral energy distribution (SED). We executed aperture photometry on archival images from the Panoramic Survey Telescope and Rapid Response System \citep[Pan-STARRS;][]{cmm+16}, Two Micron All Sky Survey \citep[2MASS;][]{scs+06}, and ALLWISE \citep{cwc+21} surveys. We identified an elliptical aperture that captured the $i$-band extent of the galaxy in Pan-STARRS data, and masked likely foreground objects. We then used this aperture, convolved with the PSF of each survey, to measure extinction-corrected AB magnitudes in all filters used by Pan-STARRS, 2MASS, and WISE. We modeled this SED using the \texttt{Prospector} stellar population synthesis modeling code \citep{jlc+21}. We ran Prospector using recommended techniques and priors and a non-parametric star-formation history (SFH) model \citep{lcj+19}, and sampled from the posterior using emcee \citep{fhl+13}. Non-parametric models for the SFH result in less bias in both stellar-mass ($M_{*}$) and star-formation rate (SFR) estimates, because specific SFH models are not excluded a priori. We included a model for dust re-radiation in the likelihood function. The spectral energy distribution of IRAS\,02044$+$7048 is shown in Figure~\ref{fig:sed}, together with the results from the Prospector analysis, and derived maximum aposteriori probability parameters are listed in Table~\ref{tab:gal}. The SFR of 1.8\,$M_{\odot}$\,yr$^{-1}$ was derived by integrating the SFH over the past 100\,Myr. 

IRAS\,02044$+$7048, and the location of FRB\,20220319D within it, are largely consistent with previous results on FRB host galaxies and environments \citep{bsp+20,hps+20,brd21,mfs+21,bha+22}. Most FRB host galaxies have detectable ongoing star formation, and some exhibit signatures of additional nuclear ionization sources. The stellar mass of IRAS\,02044$+$7048 is entirely consistent with the distribution of masses found for both repeating and so far non-repeating FRB hosts. The potential association of FRB\,20220319D with a spiral-arm feature is consistent with the results of \citet{mfs+21} based on Hubble Space Telescope imaging of eight FRB hosts. In the absence of very long baseline interferometry, our data do not have sufficient angular resolution to determine whether or not the FRB originates from within an association of young stars \citep[e.g.,][]{tgk+21}. The location of FRB\,20220319D at an offset of $\sim85\%$ of the effective radius from the galaxy nucleus is consistent with the range of previously observed FRB offsets. 

However, the SFR of IRAS\,02044$+$7048 is rather high for its stellar mass, with respect to the observed sample of hosts of so far non-repeating FRBs \citep{bha+22}. Only one so far non-repeating FRB (20190102C, at $z=0.2912$) has a comparably high specific SFR (sSFR), of $\log {\rm sSFR}=-9.74$. Although  IRAS\,02044$+$7048 is consistent with a location on the star-forming main sequence of galaxies \citep{ssc+14}, non-repeating FRB hosts as a population are consistent with originating from below this sequence. It is possible that FRBs from more actively star-forming hosts are selected against in existing radio surveys due to propagation effects in the host ISM \citep{src+21}, in which case it is not surprising that IRAS\,02044$+$7048 has a face-on orientation. 

\section{The mass of the Milky Way CGM \label{sec:5}}

The constraints on the extragalactic DM contribution along the sightline towards FRB\,20220319D are severe. Adopting the two estimates discussed above for ${\rm DM_{\rm ISM}}$ from pulsars nearby to the FRB on the sky, we find possible upper limits on the extragalactic DM contribution (${\rm DM_{\rm CGM}} + {\rm DM_{\rm IGM}} + {\rm DM_{\rm host}}$) of either 45.7\,pc\,cm$^{-3}$ or 64.3\,pc\,cm$^{-3}$. Even with our conservative estimates for ${\rm DM_{\rm ISM}}$, FRB\,20220319D has the lowest extragalactic DM yet measured for an FRB localized to a host galaxy. We thus have the opportunity to stringently bound the heretofore poorly constrained value of ${\rm DM_{\rm CGM}}$, and thus directly bound the mass of the Galactic CGM. The IGM likely contributes 7\,pc\,cm$^{-3}$ towards IRAS\,02044$+$7048, assuming an IGM baryon fraction of 0.7 \citep[e.g.,][]{mpm+20}. The detection of extended H${\rm \alpha}$ emission from the galaxy itself, and in particular at the location of FRB\,20220319D, indicates that the WIM component of the ISM is present, although we do not know where within the ISM column the FRB source is located.\footnote{With a measurement of scattering in the host ISM, it is possible to estimate ${\rm DM_{\rm host}}$ \citep{coc22}. However, upper limits on the temporal broadening of FRB\,20220319D of $\sim0.1$\,ms are not usefully constraining.} We simply assume a nominal value of ${\rm DM_{\rm host}}=10$\,pc\,cm$^{-3}$ for our analysis below. Thus, we find that for the FRB\,20220319D sightline, two possible upper limits on ${\rm DM_{\rm CGM}}$ are 28.7\,pc\,cm$^{-3}$ and 47.3\,pc\,cm$^{-3}$, depending on the assumption for ${\rm DM_{\rm ISM}}$. These upper limits are comparable to previous results from the closest known FRB source (FRB\,20200120E), in a globular cluster associated with M81, which provide possible upper limits on ${\rm DM_{\rm CGM}}$ of 32\,pc\,cm$^{-3}$ and 42\,pc\,cm$^{-3}$] depending on the choice of NE2001 or YMW16 for ${\rm DM_{\rm ISM}}$ \citep{kmn+22}. The constraints from FRB\,20200120E are affected by a reliance on the NE2001/YMW16 models for ${\rm DM_{\rm ISM}}$, and by stronger assumptions for the halo DM of M81. 

\begin{figure*}
    \centering
    \includegraphics[width=0.8\textwidth]{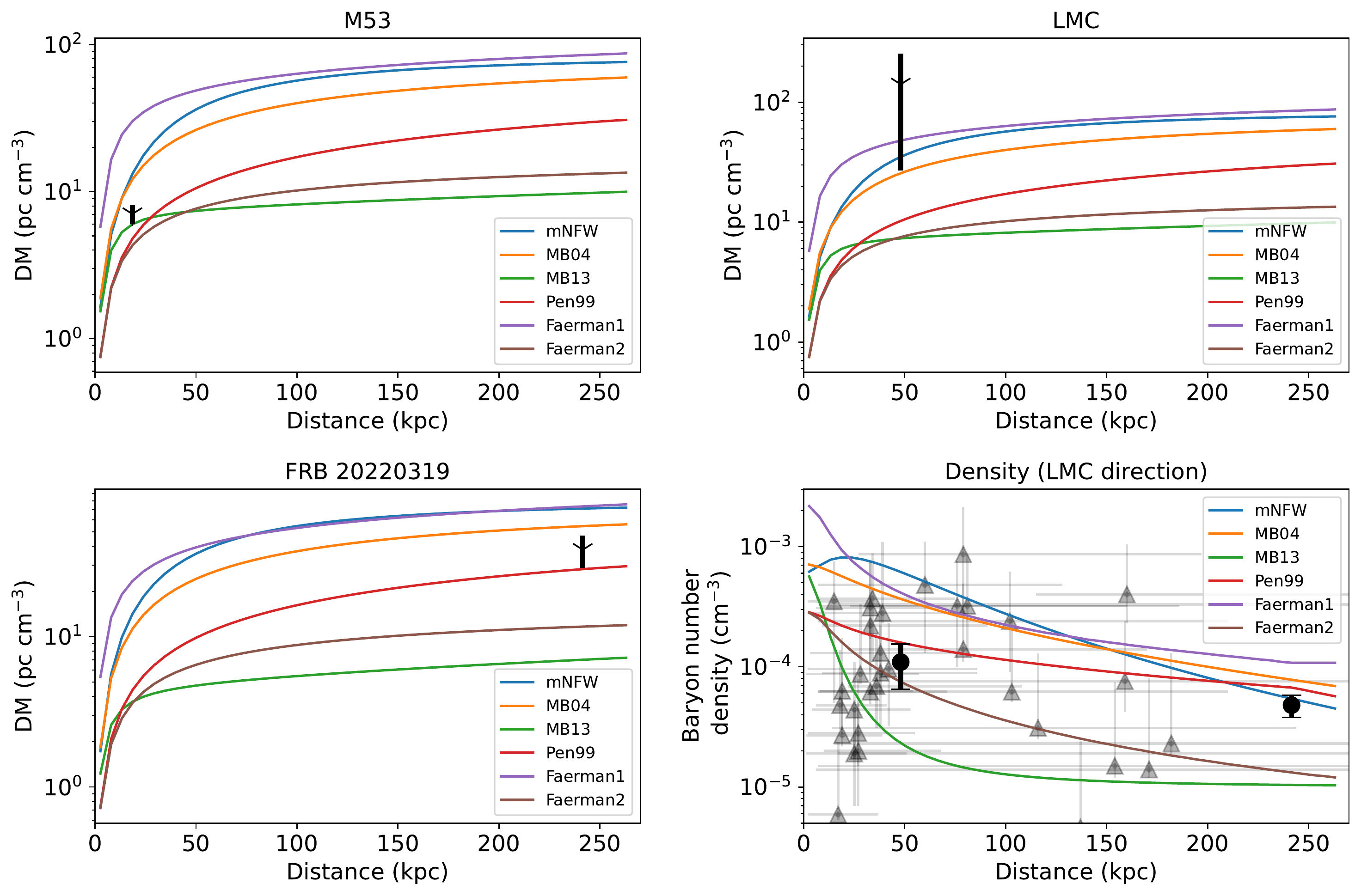}
    \caption{Illustration of the constraining power of different measurements of the CGM DM and density on a range of models. In all panels, curves show predictions of the default modified NFW (mNFW) model of \citet{pz19} (blue), the model of \citet{mb04} (orange; MB04), the model of \citet{mb13} (green; MB13), a \citet{p99} model (Pen99) assuming a core radius set to the virial radius (red), and the models of \citet{fsm17} (purple; Faerman1) and \citet{fsm20} (brown; Faerman2). \textit{Top panels:} upper limit on the CGM DM towards M53 (left) and the LMC (right). In both cases, the minimum DM (18\,pc\,cm$^{-3}$) across the sky in either the NE2001 or YMW16 models has been subtracted. The ranges indicate the range of DMs of pulsars associated with each object. \textit{Bottom left:} upper limit on the CGM DM towards FRB\,20220319D. The range indicates different assumptions for the ISM DM that correspond to the two closest pulsars on the sky. \textit{Bottom right:} constraints on the CGM density compared with the models. The dark disks indicate measurements using ram-pressure stripping modeling of the LMC \citep{sbb+15} and modeling of diffuse X-ray line emission \citep{kkk+20}. The lower limits displayed as triangles indicate estimates using ram-pressure stripping modeling of a large selection of Milky Way satellites \citep{pzp+21}.}
    \label{fig:constraints}
\end{figure*}

\begin{figure}
    \centering
    \includegraphics[width=0.48\textwidth]{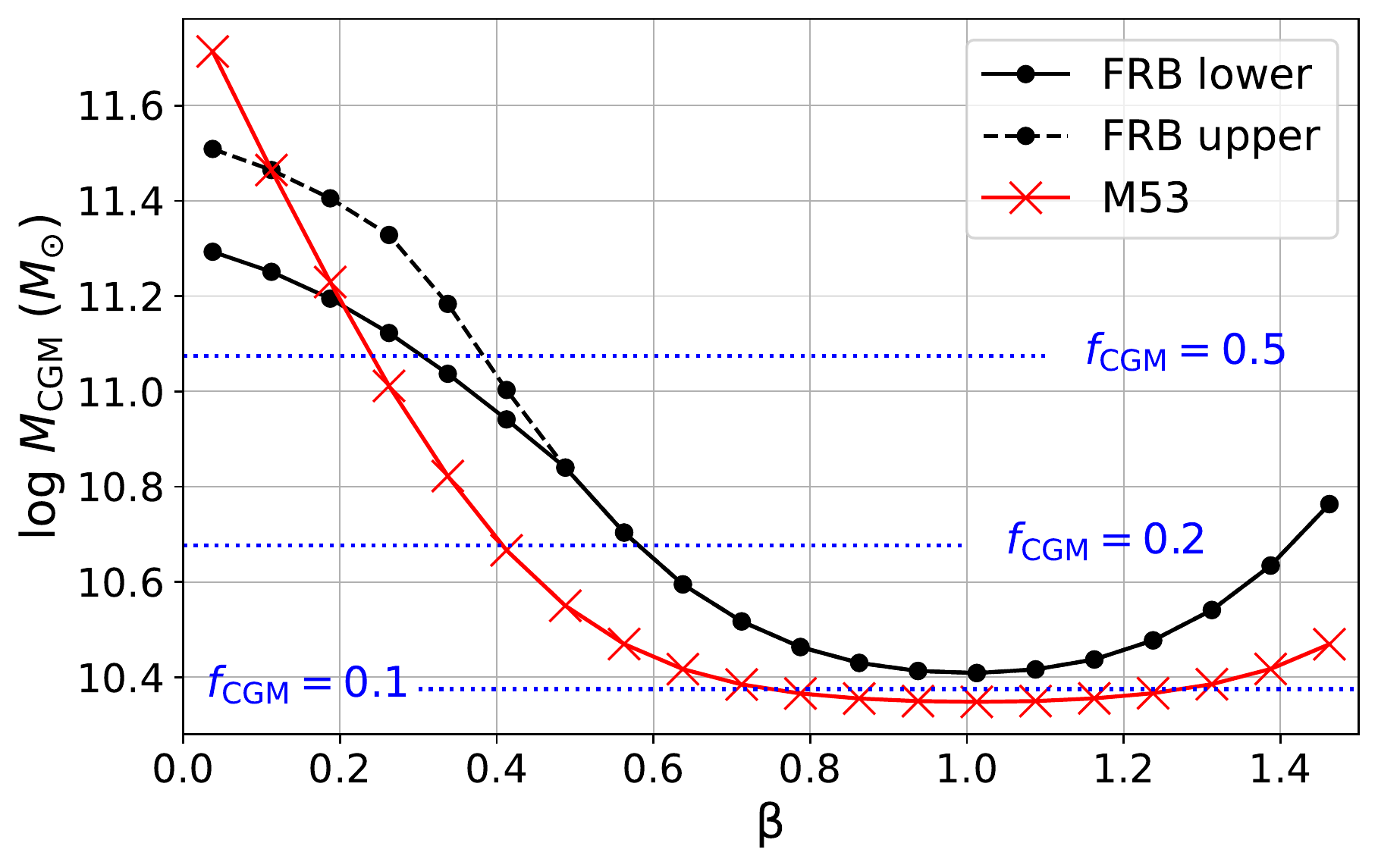}
    \caption{Upper limits on the CGM mass for different assumptions for the radial density profile. We adopt a $\beta$ model (see text for details), and show limits corresponding to the upper and lower ends of the range of constraints from FRB\,20220319D (dashed and solid lines respectively), and from the most constraining pulsar in M53 (red solid line). Typical values of $\beta$ found in the literature are $\gtrsim0.4$. For these values, the inferred $f_{\rm CGM}$ values are typically below 0.5, as indicated by blue horizontal lines. }
    \label{fig:mass}
\end{figure}

A recent synthesis of models for the Galactic CGM DM by \citet{kp20} highlighted the wide range of extant predictions. In this work, we use a representative set of models to demonstrate the implications of the constraint on ${\rm DM_{\rm CGM}}$ from FRB\,20220319D. The models are illustrated in Figure~\ref{fig:constraints}. We also consider measurements of the DM contributed by sightlines through the halo from pulsars in the LMC \citep{rcl+13} and in the distant \citep[18.5\,kpc;][]{bv21} globular cluster M53 \citep{kap+91,pqm+21}. Although the LMC is at $b\approx-33^{\circ}$ and M53 is at $b\approx+80^{\circ}$, we conservatively subtract the lowest ${\rm DM_{\rm ISM}}$ to be found in either the NE2001 or YMW16 models of 18\,pc\,cm$^{-3}$ to estimate the halo DM contributions along their sightlines. In the top two panels of Figure~\ref{fig:constraints}, we show the resulting upper limits on ${\rm DM_{\rm CGM}}$ towards M53 and the LMC, with ranges corresponding to the range of associated pulsar DMs. The bottom-left panel shows the upper limit on ${\rm DM_{\rm CGM}}$ from FRB\,20220319D, placed at the virial radius of the Milky Way halo. In all cases we show model predictions for ${\rm DM_{\rm CGM}}$ assuming a spherically symmetric galactocentric baryon halo, but with the DM evaluated along sightlines from the position of the Earth \citep{gravity19}. We evaluate the modified NFW (mNFW) and \citet{mb04} profiles following \citet{pz19}, with a total Milky Way mass of $M_{\rm tot}=1.5\times10^{12}M_{\odot}$, and a halo baryon fraction of $f_{\rm CGM}=0.75$. The \citet{p99} model is evaluated for the same $M_{\rm tot}$ and with the core radius set to the halo virial radius. The semi-empirical \citet{mb13} and \citet{fsm17} models and the semi-analytic \citet{fsm20} model are all evaluated as given. Finally, in the bottom-right panel of Figure~\ref{fig:constraints}, we also show the model predictions for the halo baryon density along the LMC sightline, together with constraints from ram-pressure stripping analyses of dwarf galaxies \citep{sbb+15,pzp+21} and modeling of OVII and OVIII emission measurements \citep{kkk+20}. 

The data are in favor of models that predict lower total CGM DMs and lower inner densities \citep[e.g.,][]{p99,fsm20,mb13}. The DMs of M53 pulsars and FRB\,20220319D deliver consistent constraints on the range of models, albeit on very different radial distance scales. The constraints from LMC pulsars are less impactful, but nonetheless also exclude the three models \citep{mb04,fsm17,pz19} considered here that predict larger values of ${\rm DM_{\rm CGM}}$. We note that our treatment of the CGM contribution to the DMs of LMC pulsars is far more conservative than that of \citet{ab10}; this is motivated by the uncertainties discussed above in estimating ${\rm DM_{\rm ISM}}$. The DM constraints are consistent with the most robust density estimates that we consider \citep{sbb+15,kkk+20}. The density estimates are however subject to the uncertainties discussed in \S\ref{sec:1}, and we proceed by considering only the constraints on ${\rm DM_{\rm CGM}}$.

We now derive constraints on the mass of the Galactic CGM, $M_{\rm CGM}$, by finding $\beta$ models that are consistent with the FRB\,20220319D and M53 DMs. When converting the electron column density to a mass column density, we assume 1.18 proton masses per electron following \citet{yt20}, which corresponds to roughly solar metallicity gas. The $\beta$ profile is widely adopted in the field to convert (column-)density estimates to halo masses \citep[e.g.,][]{mb13,sbb+15,kkk+20}, as it is empirically motivated. The electron number density at a radius $r$ is given by
\begin{equation}
n(r) = n_{0}\left[1+\left(\frac{r}{r_{c}}\right)^{2}\right]^{-3\beta/2},
\end{equation}
where $n_{0}$ is a central density, and $r_{c}$ is a core radius that we fix to 0.47\,kpc following \citet{mb13}. The exact value of the core radius is unimportant for our conclusions regarding $M_{\rm CGM}$. Typical values of $\beta$ found using various CGM tracers are in the range $0.4\lesssim\beta\lesssim0.5$ \citep[e.g.,][]{mb13,mb15,sbb+15,kkk+20}. We consider three constraints on ${\rm DM_{\rm CGM}}$: the lower and upper constraints from FRB\,20220319D, and the lower constraint from the M53 pulsars. 
For each constraint, we derive corresponding values of $M_{\rm CGM}$ from an integrated $\beta$ profile for different values of $\beta$. A useful joint constraint on $M_{\rm CGM}$ and $\beta$ is not possible with the data in hand. The results are shown in Figure~\ref{fig:mass}. 

For typical values of $\beta$ of between 0.4 and 0.5, the DM of FRB\,20220319D implies an upper limit on $\log M_{\rm CGM}$ of between 11.0 and 10.8. The M53 pulsars are even more constraining, limiting $\log M_{\rm CGM}$ to below a value of between 10.7 to 10.5. We consider the FRB\,20220319D constraints to be more conservative because the M53 estimate of ${\rm DM_{\rm CGM}}$ is very sensitive to the assumed ${\rm DM_{\rm ISM}}$ along its sightline. The mass constraints all favor values of $f_{\rm CGM}<0.5$ (recall $f_{\rm CGM}=\frac{M_{\rm CGM}\Omega_{M}}{M_{\rm tot}\Omega_{b}}$) for our fiducial value of $M_{\rm tot}=1.5\times10^{12}M_{\odot}$, with M53 potentially accommodating  values of $f_{\rm CGM}\lesssim0.2$ for reasonable values of $\beta$. 

\newpage
\section{Discussion \label{sec:6}}

Our results demonstrate that the total Galactic baryon mass is likely significantly lower than the cosmological average for a halo as massive as the Milky Way. The mass constraints we derive are  consistent with some previous observational results from analyses of OVII and OVIII emission and absorption \citep[e.g.,][]{lb17,kkk+20} and ram-pressure stripping of the LMC \citep{sbb+15}, but inconsistent with other results that posit higher CGM masses \citep[e.g.,][]{fsm17,yt20}. Analyses of the dynamics of Milky Way satellites imply a total virial mass of $M_{\rm tot}=(1.4\pm0.3)\times10^{12}M_{\odot}$ \citep{wea10}, a recent \textit{Gaia}-based analysis of the dynamics of Milky Way globular clusters yields $M_{\rm tot}=(1.3\pm0.3)\times10^{12}M_{\odot}$ \citep{ph19}, and an analysis of the dynamics of the Magellanic Stream yields $M_{\rm tot}=(1.5\pm0.3)\times10^{12}M_{\odot}$ \citep{ccb+22}. It is possible that a full accounting for the orbit of the LMC would reduce the estimates of $M_{\rm tot}$ by $\sim15\%$ \citep{cv22}. However, in all cases $f_{\rm CGM}\lesssim0.4$ is implied by FRB\,20220319D, and $f_{\rm CGM}\lesssim0.2$ is implied by the M53 DM. Our constraints are affected by different systematic effects to previous results, and do not suffer from uncertainties in modeling the chemical and thermal states of the CGM, nor from uncertainties in modeling the interaction of satellite galaxies with the CGM.  These values are consistent with several simulations of the CGM contents of galaxies like the Milky Way that account for the effects of kinetic and thermal feedback on reducing the halo baryon content. 

The observational constraints on $M_{\rm CGM}$ are affected by uncertainties in identifying the CGM DM contribution along the sightlines of interest. For the low-latitude FRB\,20220319D sightline, we adopted conservative estimates of ${\rm DM_{\rm ISM}}$, and there are uncertainties at the level of a few pc\,cm$^{-3}$ in the ${\rm DM_{\rm IGM}}$ and ${\rm DM_{\rm host}}$ terms. We therefore consider the constraints on $M_{\rm CGM}$ based on FRB\,20220319D to be robust to uncertainties in ${\rm DM_{\rm CGM}}$. The impact of uncertainty in ${\rm DM_{\rm ISM}}$ for the M53 sightline is greater given the low values of ${\rm DM_{\rm CGM}}$ under consideration. Although we attempt to be conservative, even a few-unit increase in ${\rm DM_{\rm CGM}}$ would significantly raise the derived upper limits on $M_{\rm CGM}$. Our analysis is also sensitive to how we model the distribution of baryons in the CGM. First, we do not account for sightline to sightline variance, which simulations suggest may be at the level of $\gtrsim10\%$ \citep[e.g.,][]{zpo+20,rnp22}. Second, it may be that a $\beta$ profile is not the correct model to adopt, and our constraints are sensitive to the specific value of $\beta$. Guidance on these points from simulations for future analyses of  ${\rm DM_{\rm CGM}}$ will be important. Finally, although we assume a specific helium abundance in translating the DM constraints to a total CGM mass, our results are sensitive to variations in this assumption only at the $<10\%$ level. 

The discovery of FRB\,20220319D has major implications for the search for FRBs in the local universe, and for future FRB studies of the CGM of the Milky Way and other galaxies. First, we have shown that for low Galactic latitudes ($|b|\lesssim 15^{\circ}$; see Figure~\ref{fig:globs}) there can be significant uncertainties in the widely used NE2001 and YMW16 models for ${\rm DM_{\rm ISM}}$, which may lead to nearby extragalactic FRBs being missed in surveys. In all cases we recommend that FRB surveys save candidates at DMs below the model values of ${\rm DM_{\rm ISM}}$. The possibility of missing FRBs with low extragalactic DMs also needs to be considered when using FRB samples to directly infer a characteristic ${\rm DM_{\rm CGM}}$ \citep[e.g.,][]{ppl20}. These studies are also likely to be impacted by the uncertainties in deriving ${\rm DM_{\rm CGM}}$ that in most cases will be on the order of the values of ${\rm DM_{\rm CGM}}$ allowed by our analysis (see Figure~\ref{fig:constraints}). We recommend that inferences of ${\rm DM_{\rm CGM}}$ with FRBs focus on high-latitude sightlines towards FRBs interferometrically localized to nearby galaxies, wherein uncertainties in ${\rm DM_{\rm IGM}}$ and ${\rm DM_{\rm host}}$ can be minimized. Nonetheless, if the characteristic ${\rm DM_{\rm CGM}}$ is indeed on the order of 10\,pc\,cm$^{-3}$, a very large number of local FRBs will be required to suppress variance between halo sightlines, and uncertainties in other DM contributions. These samples may be furnished by future  coherent all-sky radio monitors \citep[e.g.,][]{lll+22}. Finally, consistent with previous studies \citep{kp20,kmn+22}, our constraints on $f_{\rm CGM}$ firmly exclude the fiducial parameterizations of the widely used modified-NFW and \citet{mb04} models for the CGM DM as described by \citet{pz19}. This impacts studies that have assumed the correspondingly large values of ${\rm DM_{\rm CGM}}$ \citep[e.g.,][]{r19}, as well as forecasts for the contributions of the CGM of intervening galaxies to FRB DMs. 

The host of FRB\,20220319D, IRAS\,02044$+$7048, appears to have an unusually large specific SFR. This is despite FRB\,20220319D being the closest so far non-repeating FRB yet discovered, and the strong evolution towards higher SFRs of the star-forming main sequence of galaxies with redshift \citep{ssc+14}. Our analysis of the SFH of the  IRAS\,02044$+$7048 host galaxy suggests that most stars were formed in a burst in the last $\lesssim2$\,Gyr. Thus it is likely that the progenitor of FRB\,20220319D does not have a delay time (i.e., age from its formation epoch) in excess of this timescale. Although several progenitor scenarios remain consistent with this constraint, the additional possible association of the FRB source with a spiral arm may imply a  progenitor age that is a fraction of a $\sim100$\,Myr dynamical timescale. We note that it is not straightforward to compare our inferred SFR, which is averaged over the past 100\,Myr of the SFH, to the H${\rm \alpha}$-based SFRs common in the FRB literature \citep[e.g.,][]{bha+22}, which are sensitive to the last $\sim10$\,Myr \citep{k98}. A direct comparison of non-parametric SFH estimates may lead to a more complete view of the formation channels of FRB progenitors. 

\section{Summary and conclusions \label{sec:7}}

We present the DSA-110 discovery and interferometric localization of the nearest so far non-repeating FRB (20220319). The burst was observed with a high S/N of 79, a DM of 110.95\,pc\,cm$^{-3}$, and has a linear polarization fraction of $16\pm3\%$ and an RM of $50\pm15$\,rad\,m$^{-2}$ (Figure~\ref{fig:dynspec_pol}, and Table~\ref{tab:frb}). The FRB originated from the position (R.A. J2000, decl. J2000) = (02:08:42.7(1), $+$71:02:06.9(6)), where uncertainties in the last significant figures are given in parentheses (Figure~\ref{fig:loc}). We associate FRB\,20220319D with a spiral-arm feature of a face-on star-forming galaxy at a distance of 50\,Mpc (IRAS\,02044$+$7048; Figures \ref{fig:image} and \ref{fig:spec}; Table~\ref{tab:gal}). IRAS\,02044$+$7048 is moderately massive ($\log M_{*}=9.93\pm0.07$), with an approximately solar metallicity and an SFR averaged over the past 100\,Myr of $1.8\pm0.7\,M_{\odot}$\,yr$^{-1}$ (Figure~\ref{fig:sed}).  The low DM of FRB\,20220319D motivates a sensitive new constraint on the baryon mass of the Milky Way CGM. Our conclusions are summarized as follows.

\begin{itemize}

    \item The spatial coincidence of FRB\,20220319D and IRAS\,02044$+$7048 (false alarm probability of $<10^{-4}$), and a potential RM detection for FRB\,20220319D that is in excess of the Galactic foreground, both imply a secure association. However, the DM of FRB\,20220319D is somewhat lower than the predicted Galactic ISM DM along its sightline from leading models for the WIM distribution \citep[NE2001 and YMW16;][]{cl02,ymw17}. We show through an analysis of the DMs towards pulsar-hosting globular clusters with \textit{Gaia} parallax distances (Figure~\ref{fig:globs}) that there are significant uncertainties in NE2001 predictions for ${\rm DM_{\rm ISM}}$ for $|b|\lesssim15^{\circ}$, and that YMW16 is particularly uncertain at low latitudes in the second quadrant, where FRB\,20220319D is located (see also Table~\ref{tab:psr}). We find that estimates for ${\rm DM_{\rm ISM}}$ based on H${\rm \alpha}$ EMs perform comparably well to the models. We conclude that models for ${\rm DM_{\rm ISM}}$ towards FRB\,20220319D can accommodate an extragalactic origin, and urge FRB surveys to consider candidate bursts at DMs below standard model predictions for ${\rm DM_{\rm ISM}}$.

    \item The SFH of IRAS\,02044$+$7048 is consistent with most stars being formed in a burst within the past two Gigayears. This, together with the possible location of FRB\,20220319D within a spiral arm of IRAS\,02044$+$7048, suggests a moderately low delay time for the burst progenitor. Despite its nearby distance, the specific SFR of IRAS\,02044$+$7048 derived from SED modeling is larger than all but one so far non-repeating FRB, although the comparison with literature samples of host galaxies would be aided by the widespread use of non-parametric SFH analyses. 

    \item We find conservative upper limits on the Galactic CGM contribution to the DM of FRB\,20220319D of either 28.7\,pc\,cm$^{-3}$ or 47.3\,pc\,cm$^{-3}$: the two values are based on ${\rm DM_{\rm ISM}}$-estimates from the DMs of two pulsars nearby on the sky. These limits exclude some literature models \citep{mb04,fsm17,pz19} for the baryon distribution in the CGM (Figure~\ref{fig:constraints}). We find that consistent results are obtained from estimates of the CGM DM contributions towards pulsars in the LMC and in the distant (18.5\,kpc) high-latitude globular cluster M53. The M53 pulsars in particular provide interesting constraints in the inner regions of the Milky Way halo that complement the FRB constraint, although the M53 ${\rm DM_{\rm CGM}}$ estimate is sensitive to the assumed ISM contribution.

    \item We derive an upper limit on the mass of baryons in the Galactic CGM of $\log M_{\rm CGM}\lesssim10.8-11$ from FRB\,20220319D, and $\log M_{\rm CGM}\lesssim10.5-10.7$ from M53 (Figure~\ref{fig:mass}). The range corresponds to the range of assumed indices (0.4--0.5) of the $\beta$ profile for the baryon halo density distribution. For a fiducial total mass (baryons and dark matter) of the Milky Way of $1.5\times10^{12}M_{\odot}$, and an assumed baryon-disk mass of $6\times10^{10}M_{\odot}$, the Milky Way contains $\ll60\%$ of the cosmological-average baryon mass. This is consistent with a baryon census, predicted by galaxy-formation simulations, wherein these ``missing'' baryons are expelled from halos like the Milky Way into the IGM through kinetic and thermal feedback from AGN, supernovae and massive stars. Although our analysis relies on just a few sightlines, the conservative upper bounds on  $M_{\rm CGM}$ assuming spherical symmetry are robust to variations in the density distribution of halo baryons, and to assumptions on the chemical and thermal state of the halo.  
    
\end{itemize}

Studies of unseen baryons with FRB propagation signatures, like DMs, promise to transform our understanding of the distribution of baryons around and in between galaxies \citep{rbb+19}. As more FRBs like FRB\,20220319D are localized to nearby galaxies \citep[see also][]{kmn+22}, constraints on the CGM content of the Milky Way will continue to improve, and direct measurements of $M_{\rm CGM}$ may be possible. These measurements are likely to be important in assembling an in-situ picture of the processes whereby a galaxy grows from and impacts its environment.

\begin{acknowledgments}

The authors thank staff members of the Owens Valley Radio Observatory and the Caltech radio group, including Kristen Bernasconi, Stephanie Cha-Ramos, Sarah Harnach, Tom Klinefelter, Lori McGraw, Corey Posner, Andres Rizo, Michael Virgin, Scott White, and Thomas Zentmyer. Their tireless efforts were instrumental to the success of the DSA-110. The DSA-110 is supported by the National Science Foundation Mid-Scale Innovations Program in Astronomical Sciences (MSIP) under grant AST-1836018. We acknowledge use of the VLA calibrator manual and the radio fundamental catalog. Some of the data presented herein were obtained at the W. M. Keck Observatory, which is operated as a scientific partnership among the California Institute of Technology, the University of California and the National Aeronautics and Space Administration. The Observatory was made possible by the generous financial support of the W. M. Keck Foundation. 

\end{acknowledgments}

\facility{Hale} 
\software{astropy, CASA, heimdall, pypeit, Prospector, wsclean}

\end{document}